# In-orbit Performance of UVIT and First Results

Tandon S N[*a,b], Hutchings J B[c], Ghosh S K[d,e], Subramaniam A[b], Koshy G[b], Girish V[f], Kamath P U[b], Kathiravan S[b], Kumar A[b], Lancelot J P[b], Mahesh P K[b], Mohan R[b], Murthy J[b], Nagabhushana S[b], Pati A K[b], Postma J[g], N Kameswara Rao[b], Sankarasubramanian K[e], Sreekumar P[b], Sriram S[b], Stalin C S[b], Sutaria F[b], Sreedhar Y H[b], Barve I V[b], Mondal C[b], and Sahu S[b]

[a] Inter University Centre for Astronomy and Astrophysics, Ganeshkhind, Pune, 411007 India;
[b] Indian Institute of Astrophysics, Bengaluru, 560034 India; [c] Herzberg Institute of Astrophysics, 5071 West Saanich Rd.,Victoria, B.C. V9E 2E7, Canada; [d] National Centre for Radio Astrophysics, Ganeshkhind, Pune, 411007 India; [e]Tata Institute of Fundametal Research, Colaba, Mumbai, 400005 India; [f]ISAC, ISRO, Bengaluru, India; [g] Dept of Physics and Astronomy, University of Calgary, Calgary AB, Canada

Proofs to be sent to: sntandon@iucaa.in

**Running Title**: In-orbit Performance of UVIT and …

# ABSTRACT

The performance of the ultraviolet telescope (UVIT) on-board ASTROSAT is reported. The performance in orbit is also compared with estimates made from the calibrations done on the ground. The sensitivity is found to be within ~15% of the estimates, and the spatial resolution in the NUV is found to exceed significantly the design value of 1.8" and it is marginally better in the FUV.   Images obtained from UVIT are presented to illustrate the details revealed by the high spatial resolution. The potential of multi-band observations in the ultraviolet with high spatial resolution is illustrated by some results.

# 1. Introduction

The Ultra Violet Imaging Telescope (UVIT) is one of the payloads on the multi-wavelength satellite ASTROSAT, launched on September 28, 2015 by the Indian Space Research Organisation (ISRO). In addition to UVIT there are three X-ray payloads, viz Cadmium-Zinc Telluride Imager (CZTI), three Large Area X-ray Proportional Counters (LAXPCs), and a Soft X-ray focusing Telescope (SXT) which are co-aligned for simultaneous observations in a wide range of wavelengths from hard X-rays to ultraviolet, to study fluxes and time variations of sources simultaneously. The satellite also carries a Scanning Sky Monitor which is not co-aligned with the other payloads and scans the sky for X-ray transients. The beginnings of the UVIT project can be traced to the last decade of the last millennium, and initial ideas of its science aims and design are described by Pati (1999), Pati et al. (2003), and Rao (2003). It is primarily an imaging instrument which is designed to make high spatial resolution (FWHM < 1.8") images in a field of ~ 28' simultaneously in NUV (200-300 nm) and FUV (130-180 nm) wavelengths. The sensitivity in the FUV is ~ AB mag. 20 in 200 s of exposure. In addition to imaging, low resolution slit-less spectroscopy can be done in the NUV and FUV. The aspect of the satellite can drift up to 1' (with rates up to 3"/s) over the exposures, and frames with short exposures (< 35 ms) are added on the ground by application of a shift and add scheme to create the UV images; imaging in VIS (320-550 nm) is also done, primarily to derive the spacecraft drift. In addition to observations of time variations, UVIT is used to study hot stars in stellar clusters, star formation in nearby galaxies, the history of star formation at higher redshifts, and many other phenomena. In this paper we give a brief description of the development of the payload and its performance during the first six months in orbit.

The reader may like to compare the main features of UVIT with Galex (Martin et al. 2005 and Morrissey et al. 2007), Swift-UVOT (Roming et al. 2005) and XMM-OM (Mason et al. 2001) which have similarities to UVIT. We note here some main points of comparison between these four instruments: a) UVIT and Galex cover FUV and NUV bands, while Swift-UVOT and XMM-OM cover NUV and VIS, b) UVIT, Swift-UVOT, and XMM-OM have multiple filters to select narrower band, Galex does not have multiple filters, c) all the instruments provide slitless spectroscopy with low resolution, d) UVIT, Swift-UVOT, and XMM-OM have a spatial resolution of <2", while Galex has a spatial resolution of ~ 5", e) Galex has the largest field of view (1.2° diameter), while UVIT has a field of 28' diameter and the other two instruments have a field of 17'X17', f) in NUV all the instruments have comparable effective aperture, while in FUV Galex has an effective aperture about twice that of UVIT, and g) of the four instruments only UVIT and Galex provide for simultaneous observations in NUV and FUV.

This paper is structured as follows: Section 2 describes details of the instrument, Section 3 describes calibrations done on the ground to estimate performance of the instrument, Sections 4 describes the calibrations done in orbit and the estimated performance, Section 5 gives some initial results, and Section 6 gives a summary.

# 2. Instrumentation

Details of the design of the instrument have been published in Kumar et al. (2012a). We give below a brief description of the design.

## 2.1 Optics

The payload is configured as two co-aligned f/12 Ritchey-Chretien telescopes of aperture 375 mm. The mirrors for the telescopes were fabricated by LEOS, ISRO. One of the two telescopes observes in FUV and the other in NUV and VIS. For each of the three detectors a narrower band can be selected by a set of filters mounted on a wheel. The NUV and FUV channels also have gratings mounted on the filter wheels for slit-less spectroscopy with a resolution ~ 100. The optics of the two telescopes are shown in Fig. 1.

Fig. 1: Optics of the Ritchey-Chretien FUV telescope is shown on top. The primary mirror has an optical aperture of 375 mm and the final focal ratio is 12, and the plate scale is ~ 0.023 mm per arc second. The detector is on the extreme right and the filter-wheel precedes it by ~ 40 mm. The optics of the NUV-VIS telescope is shown at the bottom. The Ritchey-Chretien configuration of the mirrors is identical to that in FUV, but a dichroic filter is used to reflect the NUV and transmit the VIS. In addition to the detectors and filter wheels, a cylindrical lens is placed in front of the *filters for VIS to compensate for the astigmatism due to transmission through the dichroic filter.* The figure is taken from Kumar et al. (2012a).

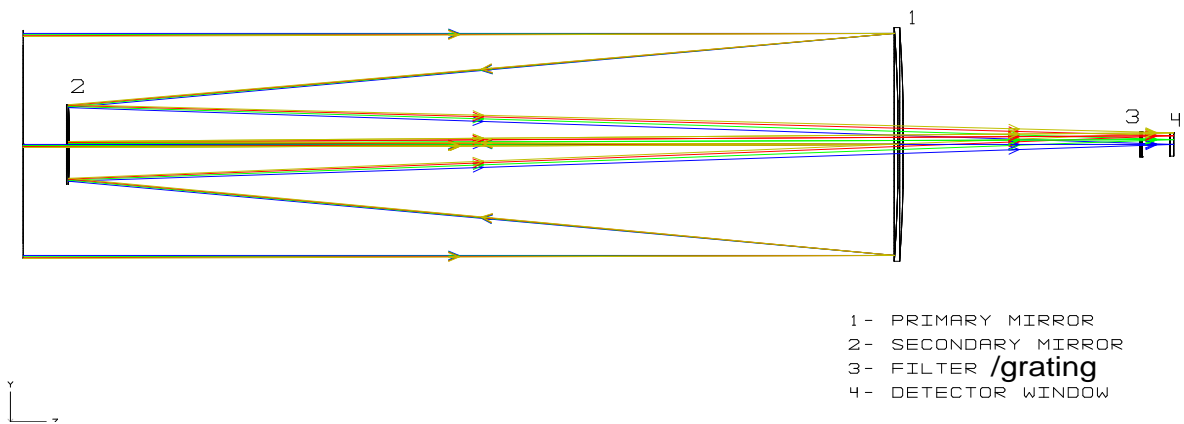

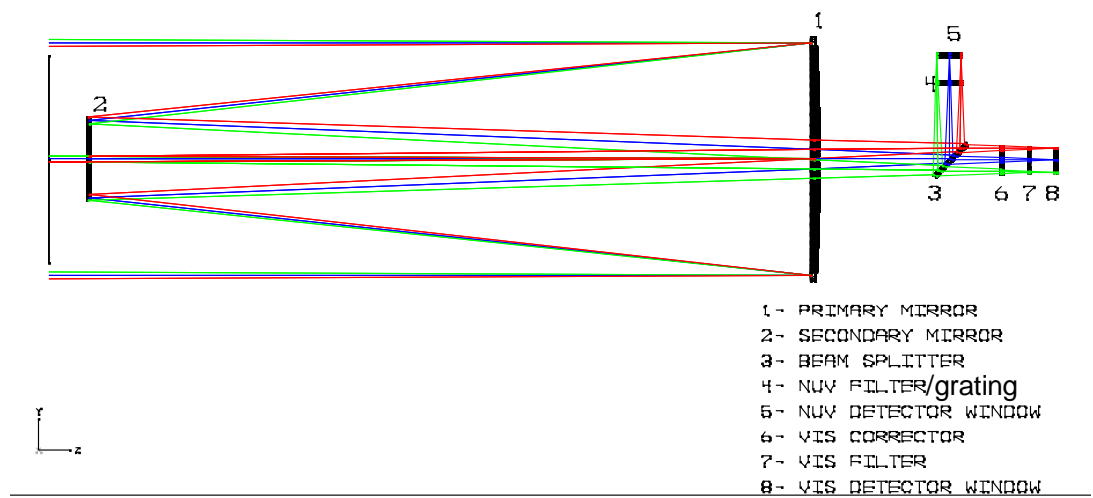

1- PRIMARY MIRROR
2- SECONDARY MIRROR
3- BEAM SPLITTER
4- NUV FILTER/grating
5- NUV DETECTOR WINDOW
6- VIS CORRECTOR
7- VIS FILTER
8- VIS DETECTOR WINDOW

## 2.2 Detectors

The three detectors are intensified C-MOS imagers which were developed for UVIT by Photek of UK, and the high voltage units for these were developed by MSSL of University College London. The schematic configuration of a detector is shown in Fig 2. The three detector systems were developed by Routes, Astroengineering, Canada under a contract with Canadian Space Agency. Each of these consists of three parts, which are called Camera Proximity Unit (which carries the detector and processing electronics), High Voltage Unit, and Electronic Unit for all the controls. All the three detector systems are identical except for the window and the photo cathode. The FUV detector has a window of MgF2 and the photo cathode is CsI. The NUV and VIS detectors have windows of Silica, and the photo cathodes for NUV and VIS are CsTe and S20 respectively. For the FUV and NUV detectors, the gaps between the photo cathodes and the micro channel plates were kept ~ 0.1 mm to minimise lateral movement of the photo-electrons and get a spatial resolution < 0.025 mm FWHM. The detectors can either work in integration mode (with a low intensification), or in photon counting mode (with a high intensification). Integration mode is used when multiple photons are expected per frame from the source. In this mode, signals in the pixels of C-MOS imager are interpreted as number of photons. Photon counting mode is used when the rate of photons from the source is less than about one per frame. In this mode, the signals above a threshold are interpreted as photon events; if a point source gives multiple photons in a frame these are counted as one. This loss of photons can be corrected for by invoking Poisson statistics as long as fraction of the frames with no photon for the source is significant, say for average rates up to ~ 2 photons /frame from the source. Further, at high photon flux (rate per second) the average signal for individual photons is reduced due to impedance of the micro channel plates. Based on the ground calibration, it is expected that for point sources the loss of photons due to this effect would be < 5% for rates up to 150 ph/s. Images can be taken with full field with a maximum frame-rate of ~ 29/s or with partial fields at rates up to ~ 600/s for a window of 100 X 100 pixels, on the C-MOS imager, corresponding to a field of ~ 5.5' X 5.5'. More details of the detectors can be found in Hutchings et al. (2007), and Postma, Hutchings & Leahy (2011).

Fig. 2: Configuration of a detector-module. The aperture is ~ 39 mm and the window is 5 mm thick on which the photo-cathode is deposited. The gap between the multi-channel-plate and the photo-cathode is 0.1 mm to 0.15 mm. The anode emits a pulse of light when the intensified pulse of electrons hits it. The light from the anode is transferred to the CMOS imager (Star250 from Fill Factory/ Cyprus) by a fibre-taper to give a linear demagnification of ~ 3. The C-MOS imager has 512 X 512 pixels each of 25 X 25 microns.

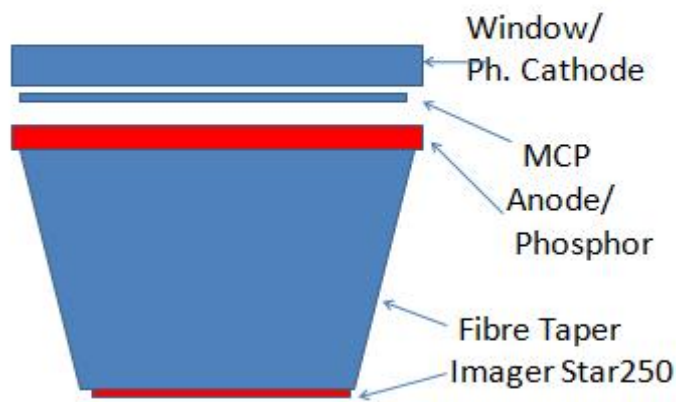

2.3 Structure

The mechanical configuration of the payload is shown in Fig.3. In addition to the supports for the mirrors, filter wheel, filters and the detectors, three baffles (see Fig. 3) are used to protect the detectors from radiation from out of field sources. The doors are also used as sun-shield by ensuring that sun is always in the plane passing through the optical axis and the normal to the doors, and the solar angle is > 45 deg. from the optical axis.

The two telescopes are integrated together by a false-cone made of Titanium alloy; this false-cone couples the payload to the spacecraft. All the elements of the telescopes which define lengths/separations along the optical axis are made of Invar to minimise defocus due to variations in the temperature, and the other parts are made in aluminium alloy. All the surfaces were black coated by an inorganic process which was developed by ISAC, ISRO for minimising contamination of the optical surfaces. In order to minimise contamination within the optical cavity, use of glues was restricted to few places. For integrating the mirrors in their frames Scotch weld epoxy adhesive 2216 B/A Gray from 3M was used. All the parts were baked in vacuum for degassing before integration. To minimise any contamination from degassing, the optical cavity was purged with clean nitrogen during testing and storage.

Fig. 3: Mechanical configuration of the payload is shown.

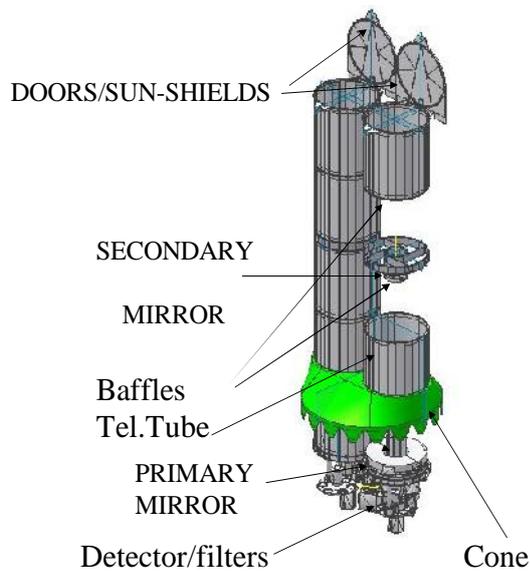

## 2.4 Contamination Control

The optical elements for the far ultraviolet can lose their efficiency by depositions of mono molecular layers. Therefore, in order to minimise contaminations the following steps were strictly observed throughout all the activities on ground for assembly and testing:

a) Using only those materials which have been certified to be free of contaminants,
b) Conducting operations in clean rooms which were monitored continuously for absence of contamination,
c) Cleaning all the incoming components/materials as per a standard protocol,
d) Before assembly, baking of all the components/parts in vacuum at high temperature
e) Purging the optical cavity by 99.999% pure N2 during the assembly and storage,
f) Limiting the number of personnel in the area of assembly and testing,
g) Cleaning of the transport containers and Thermal Vacuum Chamber etc. and checking for absence of contamination before use.

In orbit contamination can occur by initial degassing from the spacecraft, and due to polymerisation of slow and persistent residual degassing from within the payload. To avoid these, the doors of the two telescopes were kept closed for 60 days after the launch. After the doors were opened, care is taken during all the manoeuvres to avoid solar radiation falling directly in the tubes of the telescope, using the doors as sun-shield. As polymerisation is promoted by presence of ultraviolet radiation any exposure of the primary mirrors to bright earth too should be avoided, but this constraint has been given up because of requirements of slewing and pointing. To monitor effects on the sensitivity, due to contamination, an astronomical target field is observed every month in FUV. So far no significant effect has been observed over a period of 7 months (please see section 4.6 for more details).

## 3. Ground Calibrations

### 3.1 Detectors and Filters

The detectors were calibrated at the University of Calgary for the following: a) characteristics of the signal of photon events, in photon counting mode, in CMOS-imagers, b) settings of the thresholds for detection of photon events, and c) systematic errors in estimating centroids (by the on-board hardware) of the photon events and flat field effects, i.e. pixel to pixel variation in the sensitivity, for FUV and NUV detectors in photon-counting mode (Postma, Hutchings & Leahy 2011). It was found that, subject to any errors due to low frequency variations in the beam used, the flat field effects were less than 15% p-p. The errors due to low frequency variations in the beam are to be estimated from observations in orbit. The following calibrations, for the detectors, were done at IIA: a) Quantum efficiency as a function of wavelength (Stalin, Sriram & Kumar 2010), b) distortion (Girish et al. 2016); this calibration determines deviations of the scale from linearity, caused by the fibre-taper (see Fig. 2), and c) transmission of the filters as a function of wavelength.

Properties of all the filters are presented in Table 1. Based on the ground calibrations of the detectors and filters, and the reflectivity of the mirrors, effective areas can be estimated as a function of wavelength. The effective areas with all the filters, as a function of wavelength, are shown in Fig. 4.

Table 1: Properties of all the filters are shown. First column gives the name, second column gives the material or code, third column gives the average wavelength, and the last column gives the bandwidth.

| Adopted name | Filter | Wavelength (A°) | Bandwidth(A°) |
|---|---|---|---|
| FUV: | | | |
| F148W | CaF2-1 | 1480.8 | 500 |
| F148Wa | CaF2-2 | 1485.4 | 500 |
| F154W | BaF2 | 1540.8 | 380 |
| F172M | Silica | 1716.5 | 125 |
| F169M | Sapphire | 1607.7 | 290 |
| | | | |
| NUV: | | | |
| N242W | Silica-1 | 2418.2 | 785 |
| N242Wa | Silica2 | 2418.2 | 785 |
| N245M | NUVB13 | 2447.1 | 280 |
| N263M | NUVB4 | 2632.2 | 275 |
| N219M | NUVB15 | 2195.5 | 270 |
| N279N | NUVN2 | 2792.3 | 90 |
| | | | |
| VIS: | | | |
| V347M | VIS1 | 3466.2 | 400 |
| V391M | VIS2 | 3909.4 | 400 |
| V461W | VIS3 | 4614.0 | 1300 |
| V420W | BK7 | 4200.3 | 2200 |
| V435ND | ND1 | 4353.6 | 2200 |

Fig. 4    Effective areas with the various filters are shown as a function of wavelength. A common curve is shown for N242W and N242Wa.

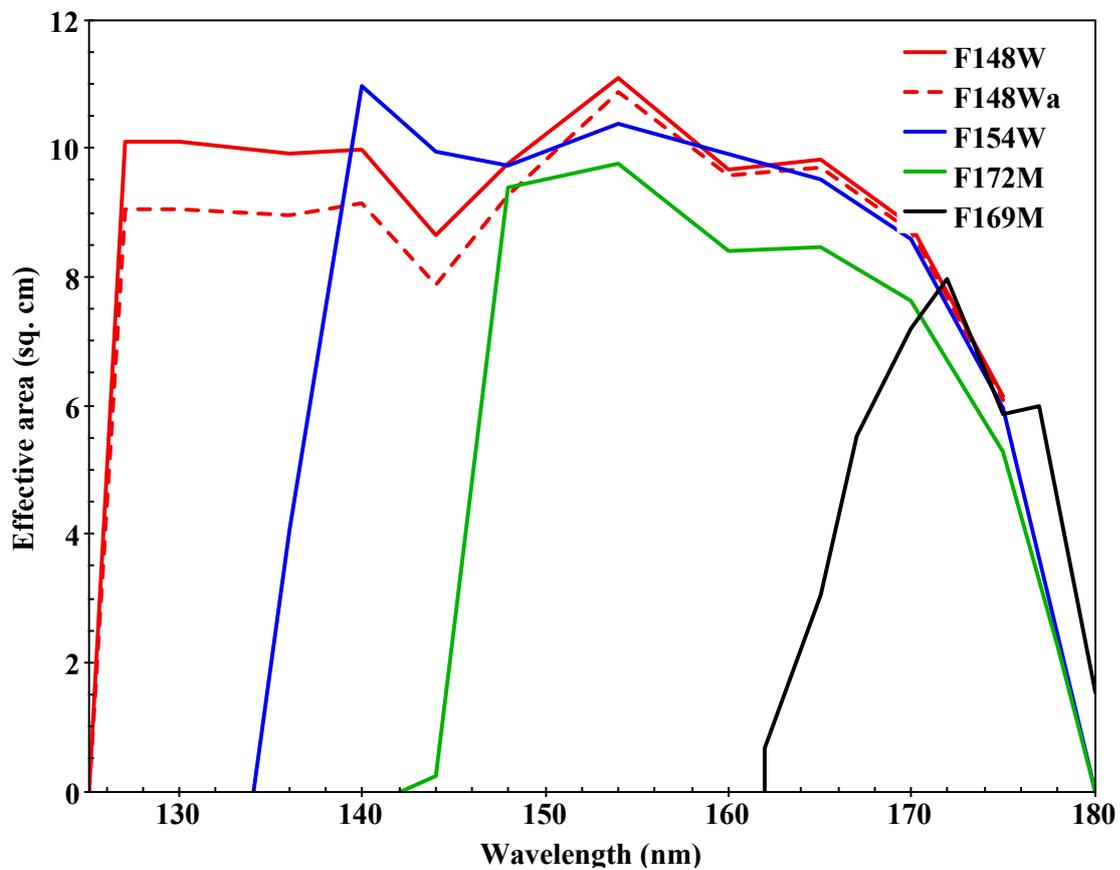

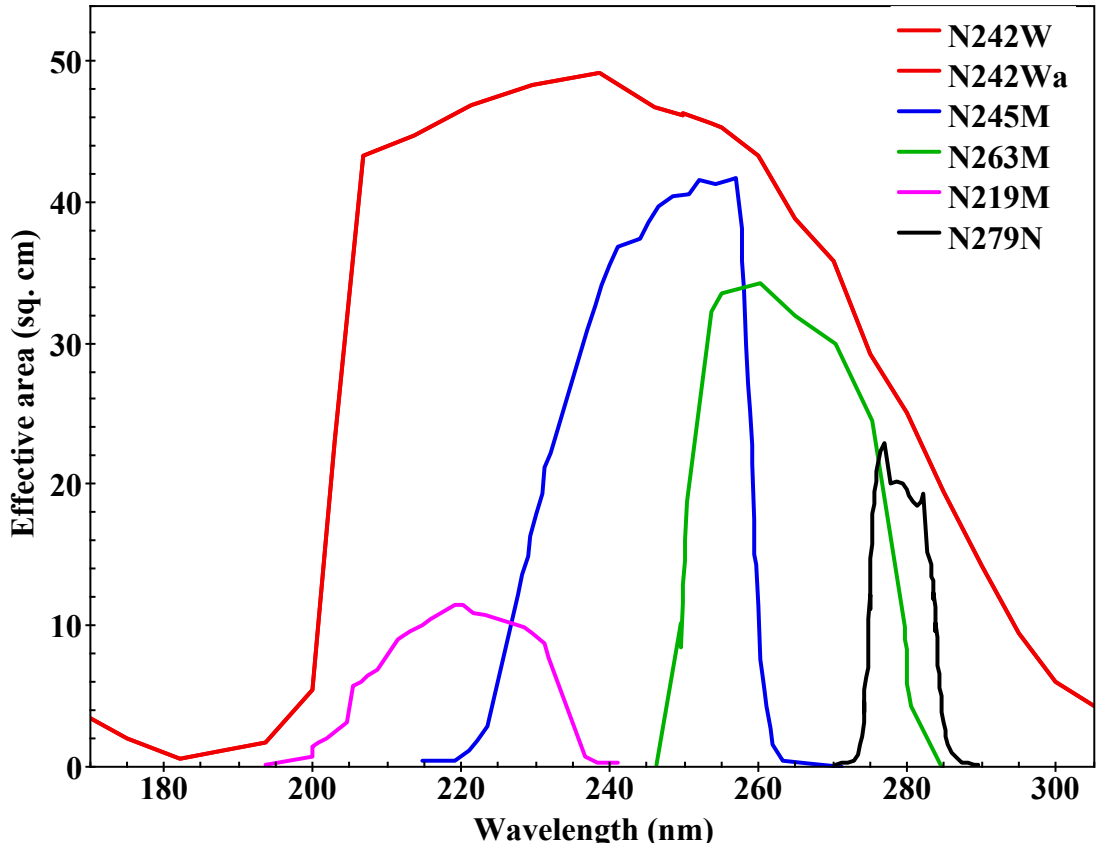

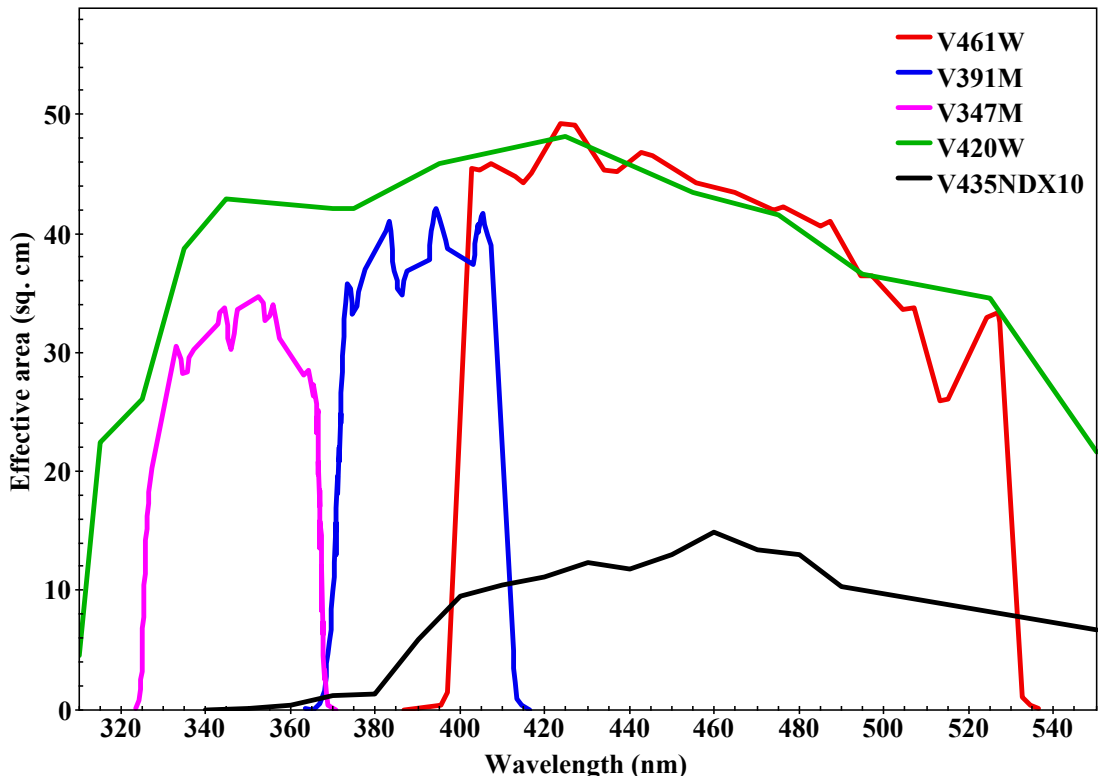

The calculated effective areas were verified for the assembled telescopes by observations with a collimated beam generated from a calibrated point source. These matched the estimates within 25%.

### 3.2 Assembled Telescopes

The assembled telescopes, with the detectors and the filters, were calibrated with a collimated beam for variation in focus with temperature, and FWHM for a point source and its variation with filter and position in the field. Typical FWHM of the image (including the effects of any aberrations in the collimated beam) was found to be ~ 1.5" for FUV and NUV in photon counting mode. The variation in focus with temperature was found to be < 0.032 mm/C. Therefore, the contribution of defocus to the PSF would be < 0.15" rms for the variability of < 3 C in the temperature expected in the orbit.

### 3.3 Assembled Payload

After assembly of the payload relative alignments of the three detectors were tested. The alignments were found to be within 50" (observations in orbit showed an angle of ~ 70" between the centres NUV and FUV fields). The payload was also tested for vibrations and for thermal effects (in vacuum) as per the standard protocol of ISRO (Kumar et al. 2012 B). After these tests the payload was integrated with the satellite and further tests were done before the launch. In order to check overall performance for imaging, simulations were done for imaging of a galaxy in FUV and NUV and the results showed that FWHM < 1.8" could be obtained (Srivastava et al. 2009).

## 4 In Orbit Calibrations and Performance

### 4.1 Electrical Tests before Opening the Doors

All functions of the payload, except imaging by the optics, were tested before opening of the doors. As high voltage generation and mechanical movements of the filter-wheels were involved, a well thought out sequence of operations, which minimised risk in case of any malfunction, was planned in advance and executed accordingly. The sequence was planned to exercise electrical activities in a gradually increasing order of complexity. A brief summary of these tests are given. Most initial operations were carried out during visibility of the spacecraft, in the eclipse part of the orbit (night time) over the primary ground station to be able to monitor and take action (if needed) swiftly.

The very first operation, after UVIT was in orbit, was to undo the contingent actions prior to launch. Prior to the launch, each of the 3 filter-wheels of UVIT were parked at respective optimal light transmitting position to minimise the risk due to failure of the motors. Filter wheels were parked to their respective light blocking configuration on the 7$^{th}$ day after launch. After 14 days in orbit, the power was turned on (one channel at a time, using 3 successive passes over ground station) and the two way communication links (commands and low bit rate telemetry over 1553B protocol) between UVIT and the Bus Management Unit (BMU) of the spacecraft were established. At every stage of testing, status of the relevant housekeeping parameters (voltages, logical states, temperatures, etc) were monitored to confirm success of the past actions before proceeding to the next phase of tests. The tests for changing the "state" of the UVIT detector modules among Low Power & Stand By were carried out next (in Low Power state the High Voltage Unit is Off, while in the Stand By state this unit is On but the high voltages are kept at near ground level). On the 15$^{th}$ day, tests for full rotations of the filter wheels were carried out by sequentially acquiring the "filter slots". Tests for the remaining states of the UVIT detector modules, viz., Active & Imaging were carried out on the 16$^{th}$ day for the VIS channel. This very first imaging in Integration Mode was carried out with high voltages programmed to very low values compared to those for nominal functioning. This activity was repeated for the other 2 channels, viz., NUV & FUV on subsequent days after confirming the good health of the individual image intensifier systems.

Next, various selectable settings of the detector system (e.g. frame read out time, window size, stacking / normalization) were exercised one channel at a time (following the sequence of VIS, then NUV and finally FUV), as per detailed plans and their satisfactory performance were validated. Each test sequence began from the 'Off' state of the detector system and ended in the 'Off' state after completion of the data collection event. During all these tests for the detectors, the corresponding filter wheels were kept at the light block configuration. These detailed tests were planned and carried out on different days due to the latency of access to the Level-1 data on ground (the Raw data received from the spacecraft undergoes several steps of processing on ground, prior to their disbursal to the payload team).

The imaging exercise was next repeated with gradually increasing values of the High Voltage settings (two 'intermediate' values @ 30% & then 60% of the 'nominal'), again for one channel at a time. The imaging data were analysed and compared with ground test data to certify their performance. These tests were carried out over 24$^{th}$ to 30$^{th}$ days post launch.

The next tests were carried out at 'nearly-nominal' values of the High Voltages (@ 75-80% of nominal) during 33$^{rd}$ to 39$^{th}$ day. Though the doors were closed, at this stage the images showed evidence for frequent Cosmic Ray Shower events, which look like a burst of photons entering the detector, and these could be used to verify the expected properties of photon-events. Next, the on board logic for triggering of the Bright Object Detect (BOD, the autonomous safety feature of the detector module against accidental exposure to bright objects), was validated. Finally, tests with 'nominal' values of high voltages were carried out (during 42$^{nd}$ to 48$^{th}$ day) in both Integration as well as Photon

Counting modes using pre-selected settings of the selectable parameters. The Cosmic Ray Shower events were used to verify properties of photon events. All electrical parameters were found to be within expected ranges.

Finally, simultaneous operation of all the three channels at nominal settings (which is expected during most of the mission life) was performed successfully on the 51$^{st}$ day post launch (November 17, 2015). At this stage the commissioning of UVIT's detector and filter wheel systems were completed and the payload was ready for tests and calibrations on the sky.

### 4.2 First Light Images

The doors were opened on the evening of 30 November 2015. After some initial glitches, images were obtained for the cluster NGC188 – selected for its high declination which avoids pointing to the ram direction and minimises bright earth exposures, and allows observation throughout the year. A raw image (as recorded in the orbit) in the VIS-channel (exposure ~ 1 s) is shown in Fig. 5. Individual images in the UV have an exposure of ~ 34 ms and are too faint to be displayed, and a processed image in NUV, for a long exposure, is shown in Fig. 6.

Fig. 5: VIS-Raw image of NGC188. The gradient, from left to right, in the background is due to change in the bias.

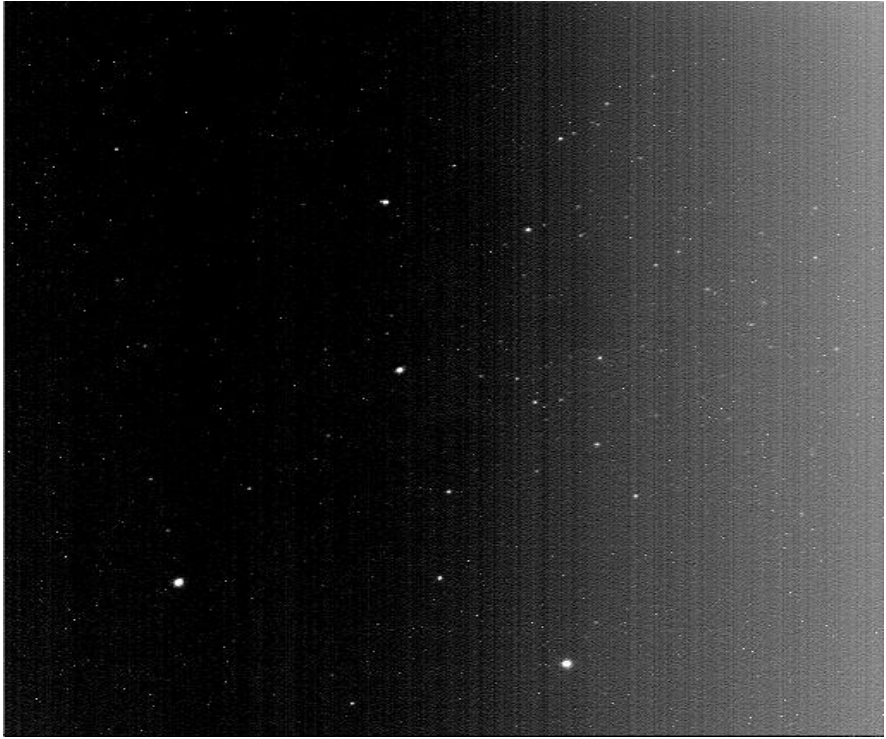

Fig. 6: A processed NUV image of NGC188. The orientation is arbitrary.

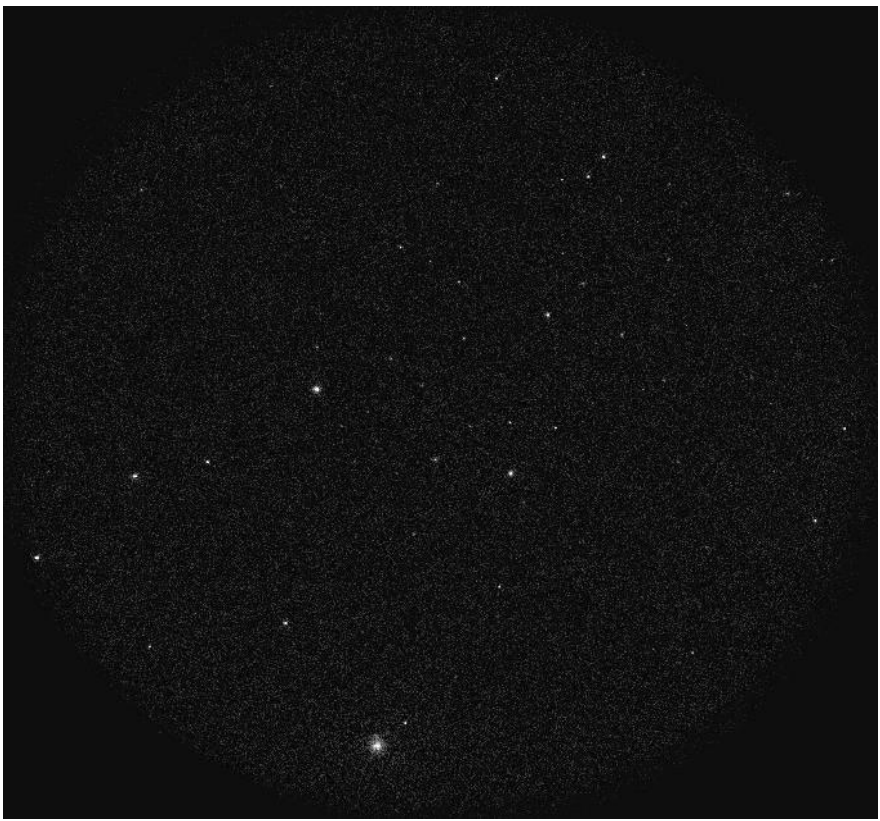

### 4.3 Sensitivity

A primary standard (White-dwarf Hz4) was used to calibrate the sensitivity, in the various ultraviolet filters, near the centre of the field. The magnitude system adopted for UVIT is the AB magnitude system (Oke 1974), and hence the zero-point magnitudes derived are in this system. The results shown in Table 2 are based on the spectrum of Hz4 from the CALSPEC database at STScI for NUV and FUV. The entries in this table match the estimates made from calibrations on the ground within ~ 15% (see Subramaniam et al. 2016 for more details).

In the NUV the calibration was only done for 2 of the 6 filters, as the count rate for the other filters was estimated (for Hz4) as beyond the acceptable limit. Calibration for these would be done in the future with some other standard.

Table 2: Sensitivity for the different filters is shown. The calibration is done with the primary standard Hz4. The columns are as follows: i) The filter, ii) the mean wavelength (A), **which** is the weighted mean based on the measured effective areas on the ground, iii) the bandwidth (A) between two half-effective-area wavelengths, iv) mean effective area referred to the mean wavelength and the bandwidth, v) "Unit Conversion", i.e. the flux (ergs/(second. sq cm**.** A)) at the mean wavelength which gives one count per second for the spectral shape of Hz4, and vi) "Zero Point AB-mag." corresponding to the "Unit Conversion". For comparision the parameters given by Morrissey et al. (2007) for Galex are also listed. All the entries in this table are from Subramaniam et al. (2016).

| Filter | λ(mean) | Δλ | EEA | Unit conv | Zero Point |
|---|---|---|---|---|---|
| **FUV** | | | | | |
| CaF2 -1 | 1480.8 | 500 | 9.21 | 0.29153E-14 | 18.08 |
| BaF2 | 1540.8 | 380 | 9.83 | 0.26921E-14 | 17.80 |
| Sapphire | 1607.7 | 290 | 10.08 | 0.42305E-14 | 17.50 |
| Silica | 1716.5 | 125 | 8.96 | 0.18696E-13 | 16.38 |
| **NUV** | | | | | |
| NUVB15 | 2195.5 | 270 | 12.31 | 0.54151E-14 | 16.55 |
| NUVN2 | 2792.3 | 90 | 24.55 | 0.36443E-14 | 16.46 |
| **GALEX** | | | | | |
| FUV | 1538.6 | | 19.6 | 0.140E-14 | 18.82 |
| NUV | 2315.7 | | 33.6 | 0.206E-15 | 20.08 |

4.4 Spatial Resolution and Astrometry

Pointing of the satellite has slow drifts up to +- 1' with a rate up to 3"/s. Therefore, in the ultraviolet channels images are recorded at a rate ~ 29/s or higher to keep the blur in individual images to be << 1". The final images are made on the ground by shift and add algorithm. As many ultraviolet images have very low flux it is not possible to derive drift from the images taken at 29/s. In many fields, the NUV images can be used to find the drift by binning the images every one second. A more secure solution to find drift is to use images taken by the visible channel every ~ 1 s. The final spatial resolution depends on the intrinsic quality of the images given by the optics and detector combination and any errors in correcting for the drift. The final FWHM of the images is found to be ~ 1.2" for NUV and ~ 1.5" for FUV – these are significantly better the goal of 1.8". Typical radial profiles for the PSF in FUV and NUV are shown in Fig. 7. A part of the image, centred on NGC2336, in NUV and FUV is shown in Fig. 8 to illustrate the spatial resolution. Fig. 9 shows the errors in relative positions of many stars. The top panel shows errors in the NUV image of NGC2336 with reference to a standard ground based image, and the bottom panel shows errors in the FUV image of NGC1851 with reference to the FUV-image from Galex. It is seen that the errors are <1", except for a few deviant stars.

Fig. 7: Typical radial profiles for the stellar images in NUV (top) and FUV (bottom) are shown. The X-axis is in units of 1/8 pixel or ~ 0.41". The profile for FUV shows a significant pedestal extending up to a radius of ~ 3".

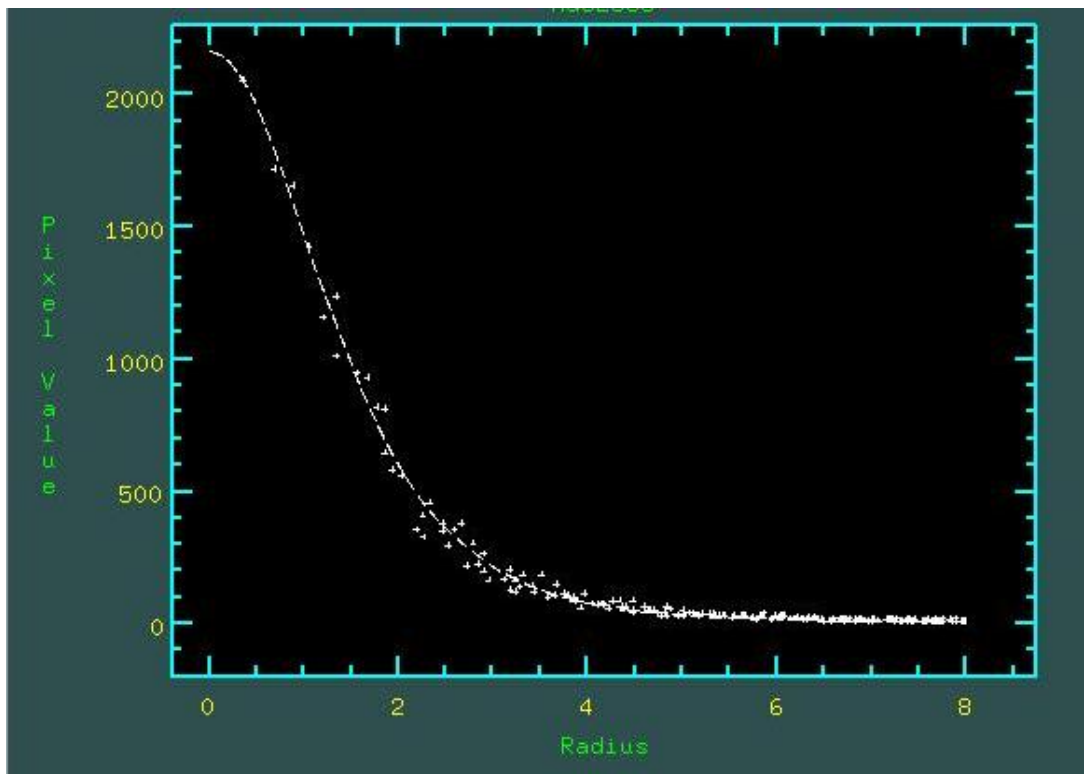

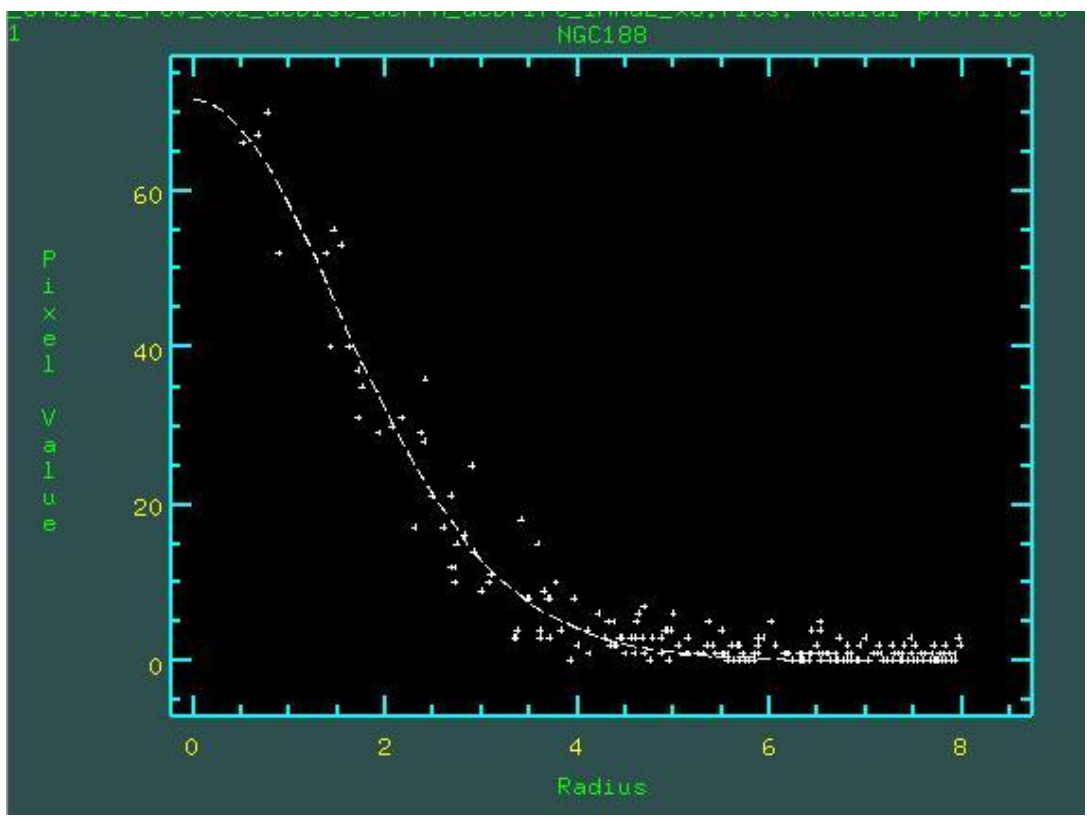

Fig. 8: From top to bottom: i) Image of NGC2336 in NUV in photon counting mode. The arms of the small galaxy on the left, separated by ~ 3", are resolved; total width of the frame is ~ 7', ii) Image of NGC 2336 in FUV .

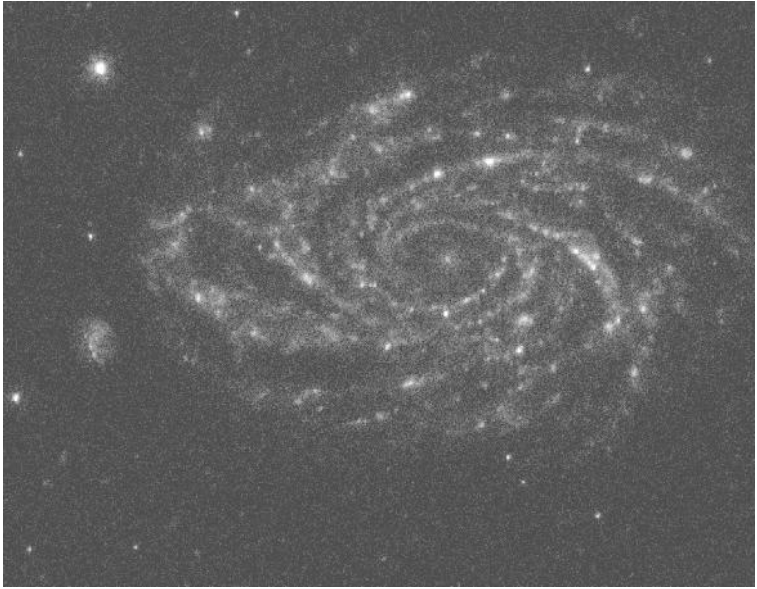

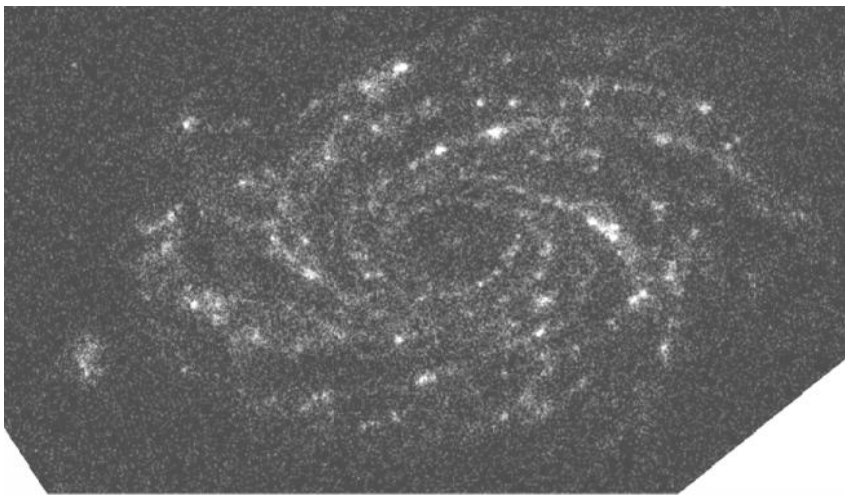

Fig. 9 : Relative astrometric errors are shown as vectors for an NUV image of NGC2336, and an FUV image of NGC1851 after finding the best fit to the stars in the corresponding DSS2 (Blue) image and Galex (FUV) image respectively. The smallest and the largrst vectors are shown in black and blue respectively, and the rest are shown in red.Tails of the vectors are at positions of the stars. The horizontal vectors at the bottom show magnitude scale of the vectors. Each sub-pixel is ~ 0.41".

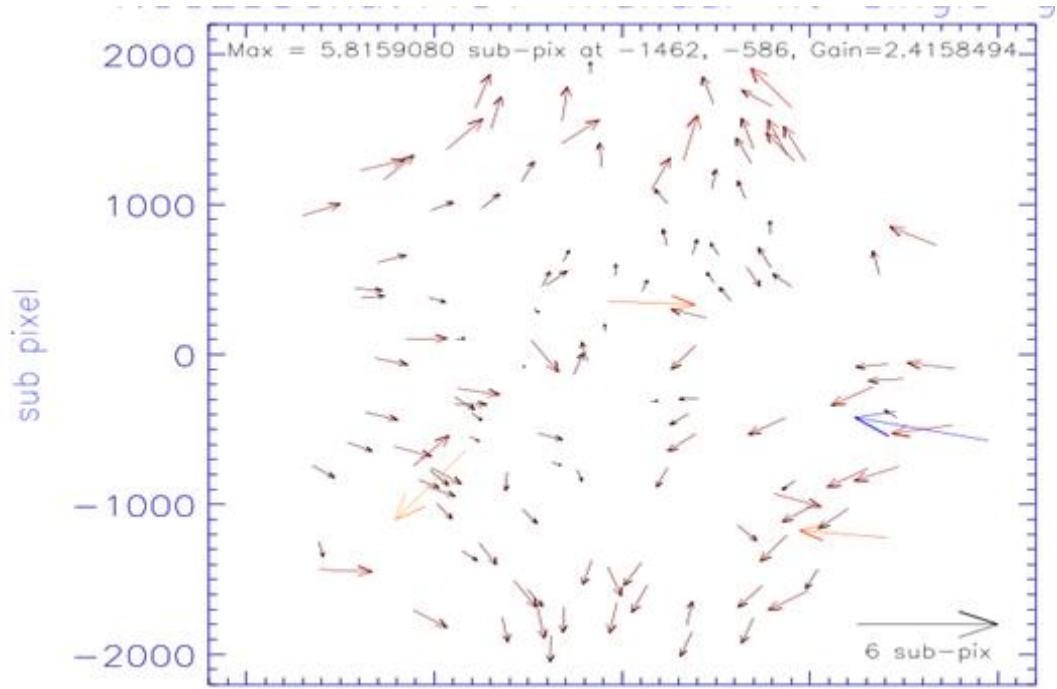

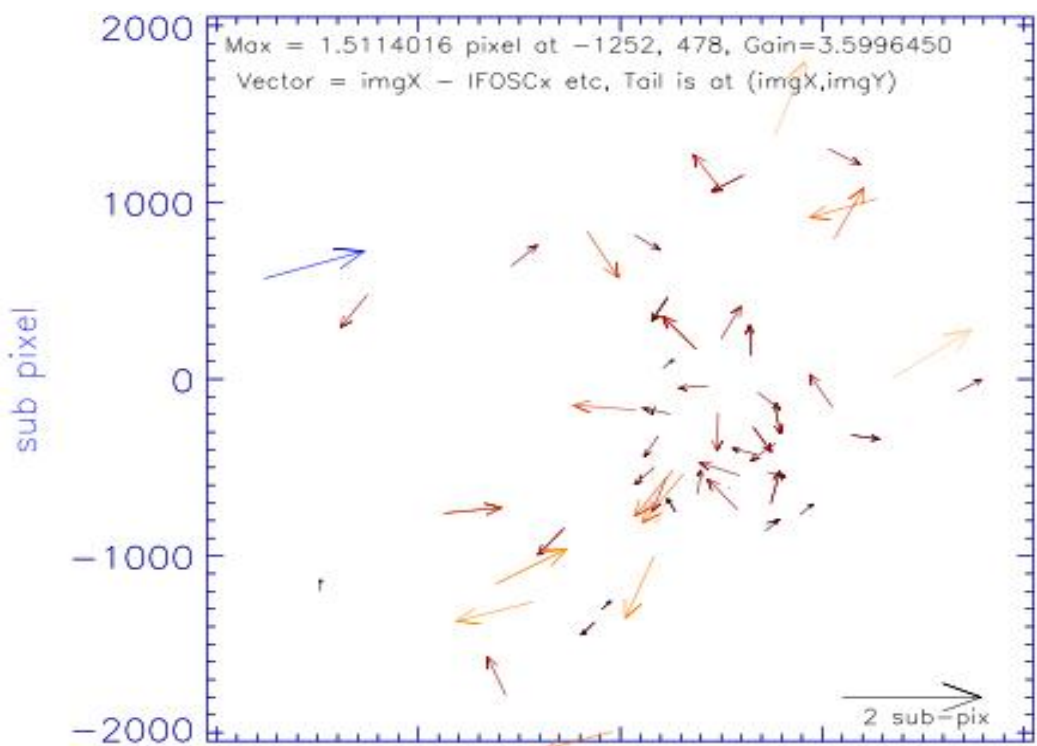

### 4.5 Spectral Calibration

Calibration for sensitivity of the gratings was done with Hz4 (a primary photometric standard), and the dispersions were calibrated with the planetary nebula NGC40. The effective areas as a function of wavelength are shown in Fig. 10. An image of the NUV spectrum is shown in Fig. 11, and one dimensional plots of the signals obtained for NGC40 are shown along the dispersion axis in Fig. 12. Only one of the two gratings in FUV has been calibrated. The present results for NUV are adversely effected by saturation and are only indicative of the effective areas and spectral resolution. More calibrations would be done in NUV with elimination of saturation, and the second FUV grating would be calibrated. The spectral resolution, for NUV (first order) and FUV (second order), is ~ 100.

Fig. 10: Effective areas (in square cm on the Y-axis) as a function of wavelength (Angstrom along the X-axis) are shown for the gratings.  A. Left panel shows effective area in the first order for the NUV-grating. The smooth blue curve shows the results from ground calibrations.  The three wavy curves are the results from three different exposures of Hz4; the wavy nature is due to effects of saturation. B.  Right panel shows effective area in the second order for the FUV-grating. The red curve shows the results from ground calibrations. The blue curve shows the results from observations of Hz4.

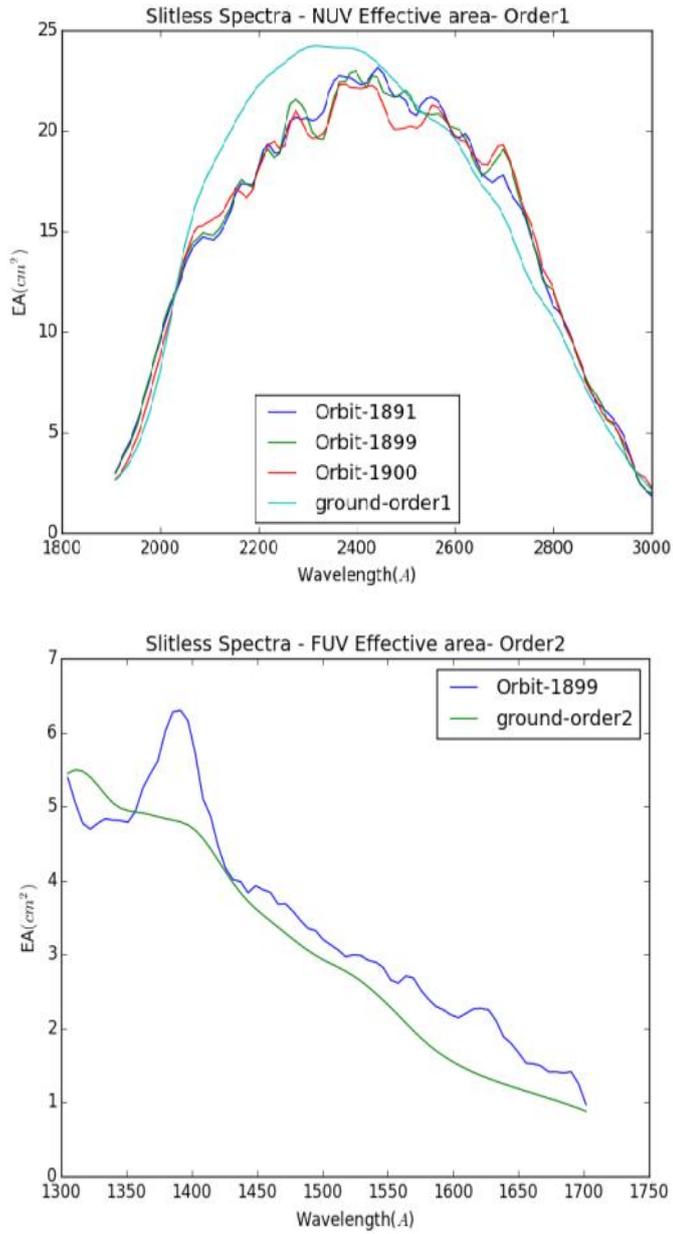

Fig. 11: Image of the NUV spectrum for NGC140 is shown. Spectrum for NGC140 is on the right of the centre. The sharp spot is the "zero order", while the first, second, and third orders are seen in the lower part. The "negative orders" are seen on top of the "zero order" and only carry a minor part of the energy.

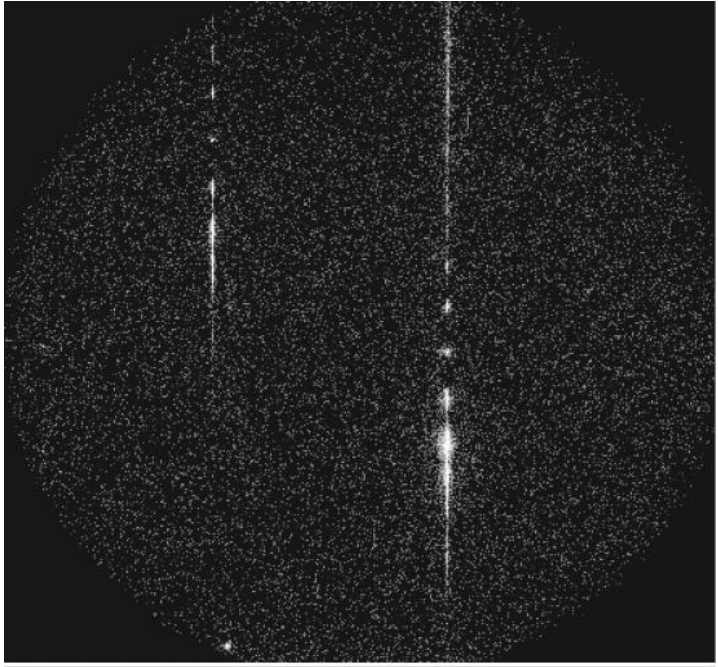

Fig. 12: Signals are shown along the dispersion axis for NGC40.  A. Top-left panel shows the signal (Y-axis) vs position (X-axis in unit of ¼ pixel) for NUV. The zero order is seen as a sharp spike near "1100" on X-axis, on the left of it are seen the main first and second orders, and on the right are seen the negative first and second orders. B. Top-right panel shows the signal vs wavelength (in Angstroms) in the first order of NUV, after fitting positions of the Zero-order and prominent lines at 2297 A and 2527 A. Spiky nature of the signal is due to saturation. C. Bottom-left panel shows the signal (Y-axis) vs position (X-axis in unit of ¼ pixel) for FUV. The zero order is seen as a sharp spike near "900" on X-axis, on the left of it are seen the main first and second orders, and on the right are seen the negative first and second orders. D. Bottom-right panel shows the signal vs wavelength (in Angstroms) in the second order of FUV, after fitting positions of the Zero-order and prominent lines at 1405 A, 1556 A, 1641 A, and 1723 A .

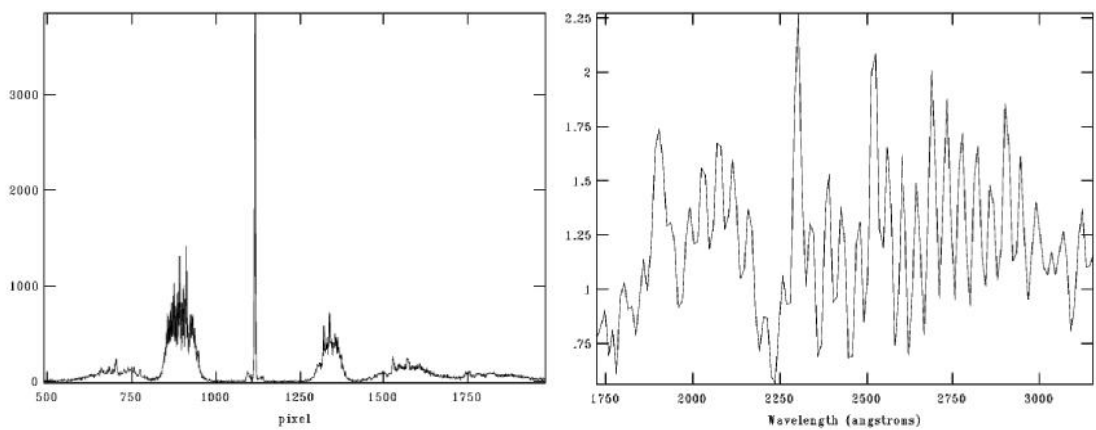

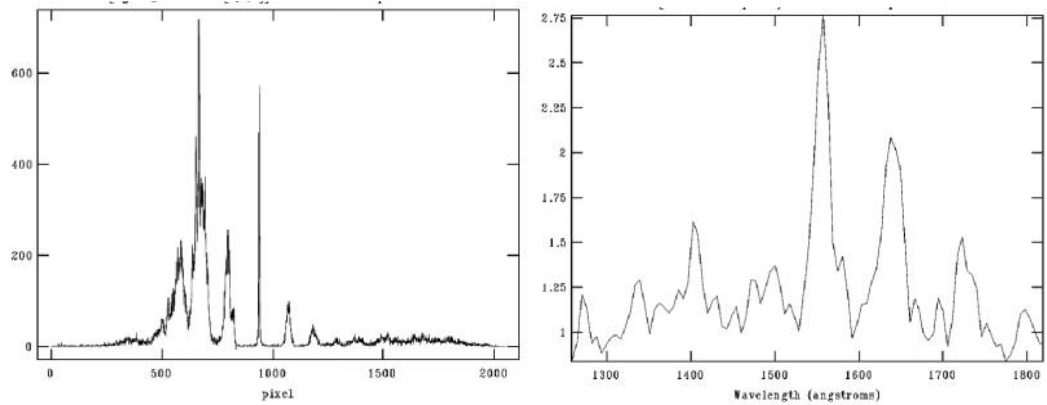

### 4.6 Sensitivity Variation

Transmission of the optics and quantum efficiency of the detectors can change with time. In particular, small contamination can lead to reduction in transmission of the FUV optics. Therefore, two bright stars in NGC188 are repeatedly observed every month to get a measure of any variation in sensitivity. So far all the results indicate that any reduction in sensitivity of FUV, with CaF2 filter, is less than 5% over a period of ~ 7 months in the orbit. This gives confidence that any contamination, if at all, of the optics is not occurring at a high rate.

### 4.7 Background

The background in FUV and NUV arises from geo-coronal lines, zodiacal light, and Galaxy. The contribution of geo-coronal lines vary with time of the day and with the level of solar activity. The other contributions depend on the direction of view. In addition to these a significant contribution is found from the cosmic ray interactions ( seen as bunches of large number of photons in about 3 frames every second) which give a background ~ 120 c/s. In the FUV it can be the major background in dark fields. This background can be minimised by rejecting the frames with number of photons more than a threshold, e.g. average number observed per frame plus 3-sigma.

.

### 4.8 Artifacts in the Images

The optics are not expected to give any ghosts which are outside the PSF of point sources. However, we do see faint streaks in some NUV images which are inferred as due to some star near the edge of the field. With more observations we expect to develop a better understanding of these streaks. The effect of saturation on the PSF is illustrated in Fig. 13.

Fig. 13: Effect of saturation on the PSF is shown for a NUV image; the rate of photons is >1/frame. The central peak is surrounded by a dip at a diameter ~ 3 pixels of the CMOS-imager, and is related to the algorithm for detecting photon events (see Srivastava et al. 2009 for more details).

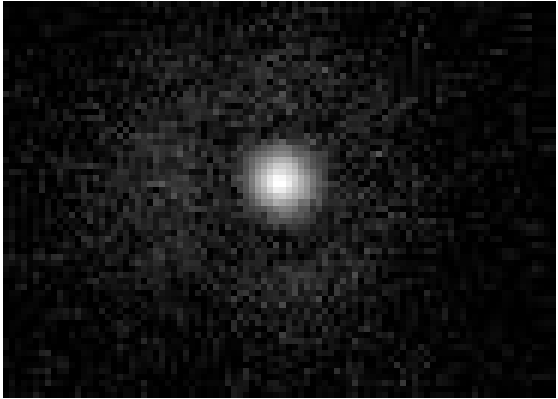

## 5   Some Results

UVIT has several filters in the FUV and NUV channels which can be used to sample the spectral energy distribution of the observed source. This capability can be used to estimate accurate spectral properties of the sources which have energy generation due to different physical processes or consist of multiple components. Further, by virtue of a much better spatial resolution it can reveal more details of the crowded fields as compared to what could be done with Galex. We present here two preliminary results to demonstrate these capabilities.

a) The field of the old open cluster, NGC 188, has two UV bright stars, which are relatively faint in the visual bands. These stars are thought to belong to the class of stars known as sub-dwarfs. The UVIT measurements of fluxes of one of these sub-dwarfs in NGC 188 are shown in Fig. 14. The other measurements in UV from Galex and UIT (Stecher et al. 1997), and the measurements in visual and near-IR from ground observations are also shown. The flux estimations from UVIT compare well with those from UIT and Galex. The figure demonstrates that the filter systems in UVIT help to widely sample the flux distribution in the UV. Note that the figure shows fluxes in the FUV from all the filters, and fluxes in the NUV only from two filters. These flux distributions can be used to compare with models to estimate the temperature, mass and radius of the source, which is known to be a single lined binary (Geller et al. 2008), with a suspected hot companion (Landsman et al. 1998).

Fig. 14: The spectral energy distribution of a sub-dwarf star (S1-WOCS5885) in the open cluster NGC 188. The fluxes obtained from the UVIT observations are shown as green points. The optical data are taken from Sarajedini et al. (1999) and the near-IR data are from the 2MASS catalogue. The figure is taken from Subramaniam et al. (2016).

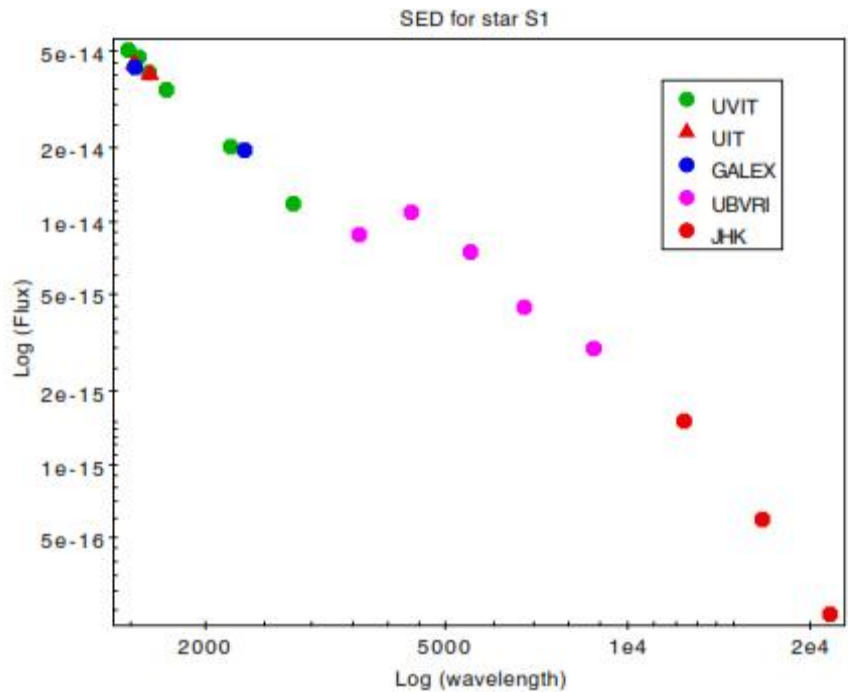

b) The globular cluster, NGC 1851 has many UV-bright stars in its core and has been observed by Galex. The central part of its image obtained by UVIT in the FUV, using the CaF2 filter, is compared with the FUV images from Galex and HST in Fig. 15. The figure clearly demonstrates that UVIT resolves many more hot stars than Galex could, and thus provides a much richer sample of stars to understand properties of the UV bright population.

Fig. 15 : The FUV image of NGC 1851 from UVIT (Caf2 filter) is shown along with the images from the GALEX and HST missions. The arrows are shown to guide the reader to the pattern of stars which are resolved by UVIT.

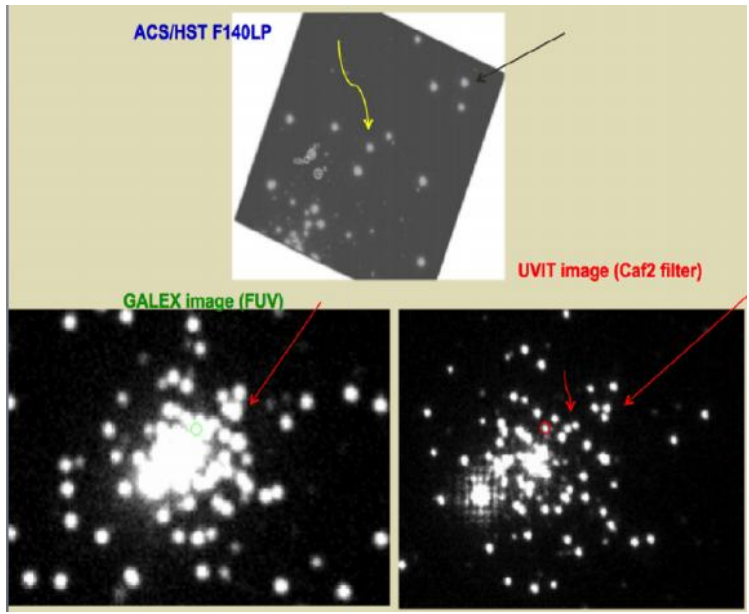

The UV-colour-magnitude diagram (CMD) for the bright stars, using the FUV magnitudes in CaF2 filter and NUV magnitudes in N2 filter, is shown in Fig. 16. In order to identify various evolutionary sequences, the isochrone generated for the UVIT filters, using the Flexible Stellar Population Synthesis (FSPS) model by Conroy, Gunn & White (2009), is also plotted. The isochrone is derived for an age of 10Gyr, and the Horizontal Branch (HB) and White Dwarf (WD) sequences are identified in the figure. The isochrone is corrected for the distance to NGC 1851 but is not reddening corrected, which is the reason for the isochrone appearing slightly brighter than the observed magnitudes. The CMD shows only the HB which appears as a tilted sequence with a large range in temperature, and is similar to the HB in CMD of NGC1851 derived from Galex (Schiavon et al. 2012). We also detect some stars brighter than the HB stars, which could be candidate binaries. Detailed analysis of the HB stars is in progress.

Fig. 16: The Colour-magnitude diagram of NGC 1851 is shown along with the isochrones derived for the UVIT filters using FSPS models. The axes are shown in AB magnitudes, and the observed stars are shown as red dots. The Horizontal Branch (HB) sequence and the White Dwarf (WD) sequence are marked.

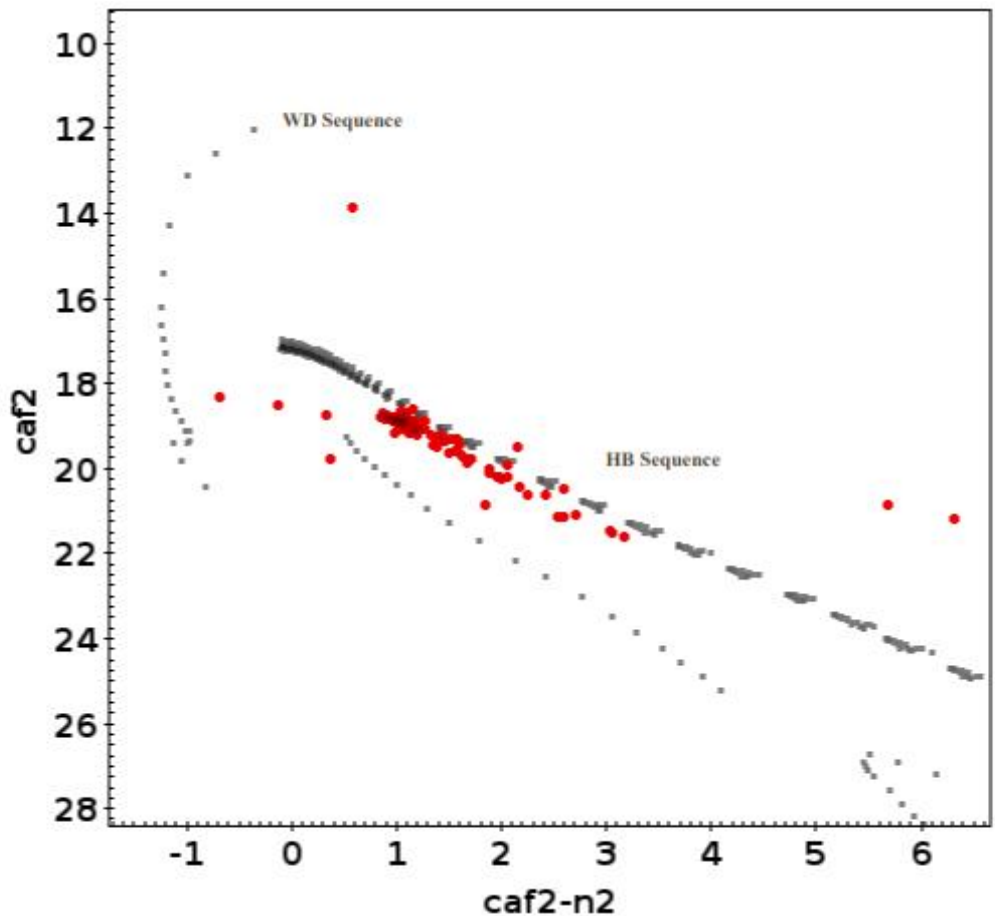

## 6  Summary

The performance of UVIT in the orbit has been described.  The initial calibrations and observations show that the payload is giving excellent and stable performance in orbit. In particular the spatial resolution in the NUV has significantly exceeded specifications. The sensitivity, as represented by zero magnitudes in the various bands is lower by up to 15% compared with the estimates made from ground calibrations. Some preliminary results are presented to illustrate the potential of multiple UV-band observations with the high spatial resolution of UVIT.

**Acknowledgement**

contributed to the work presented in this paper. Several groups from ISAC (ISRO), Bengaluru, and IISU (ISRO), Trivandrum have contributed to the design, fabrication, and testing of the payload. The Mission Group (ISAC) and ISTRSAC (ISAC) continue to provide support in making observations with, and reception and initial processing of the data. We gratefully thank all the individuals involved in the various teams for providing their support to the project from the early stages of the design to launch and observations with it in the orbit.

# References

Conroy, C., Gunn, W., White, M., 2009, ApJ, **699**, 486

Girish V., Tandon S. N., Sriram S., Kumar A., Postma J 2016, "Mapping Distortion of Detectors in UVIT Onboard Astrosat Observatory" submitted for publication

Geller, A.M., Mathieu, R. D., Harris, H. C., McClure, R. D. 2008, AJ, **135,** 2264

Hutchings, J.B., Postma J., Asquin D., Leahy D. 2007, *PASP,* **119**,1152

Kumar, A. et al. 2012a, Proc. of SPIE, **8443**, 84431N

Kumar A. et al. 2012b, Proc. of SPIE, **8443**, 84434R

Landsman, W., Bohlin, R. C., Neff, S. G., O'Connell, R. W., Roberts, M. S., Smith, A. M., Andrew, M., Stecher, T. P. 1998, AJ, **116**, 789

Martin, D.C. et al. 2005, ApJ **619**, L1

Mason, K. O. et al. 2001, A&A, **365**, L36

Morrissey, P. et al. 2007, Ap J Suppl. Series, **173**, 682

Oke, J.B. 1974, ApJS, 27, 21

Pati, A.K. 1999, BASI,**27**, 295

Pati, A.K., Mahesh P. K., Nagabhusan, S., Subramanian, V. K. 2003, BASI, **31**, 479

Postma, J., Hutchings, J. B., Leahy, D. 2011, PASP, **123**, 833

Rao, N. K. 2003, BASI, **31**: 249

Roming, P. W. A. et al. 2005, Space Sci Rev **120**, 95

Sarajedini, A., von Hippel, T., Kozhurina-Platais, V., Demarque, P. 1999, AJ, **118**, 2894

END